\documentclass[a4paper]{article}

\usepackage{INTERSPEECH2021}
\usepackage{listings}
\usepackage{hyperref}
\usepackage{adjustbox}
\usepackage{amsmath}
\usepackage{amsthm}
\usepackage{bigdelim}

\usepackage[labelformat=simple]{subcaption}

\usepackage[font=small,labelfont=bf]{caption}
\usepackage{bbm}
\usepackage{pifont}
\usepackage{enumitem}
\usepackage{multirow}
\usepackage{adjustbox}
\usepackage{booktabs}
\usepackage[hang,flushmargin]{footmisc} 

\usepackage[T1]{fontenc}

\usepackage{pifont}
\newcommand{\cmark}{\ding{51}}%
\newcommand{\xmark}{\ding{55}}%

\usepackage{xcolor}
\usepackage[linesnumbered,ruled,vlined]{algorithm2e}
\allowdisplaybreaks

\usepackage[labelformat=simple]{subcaption}

\usepackage[font=small,labelfont=bf]{caption}

\SetCommentSty{mycommfont}

\SetKwInput{KwInput}{Input}                % Set the Input
\SetKwInput{KwOutput}{Output}              % set the Output
\SetKwInput{KwInit}{Init.}              % set the Initialization

\definecolor{mygreen}{rgb}{0,0.6,0}
\definecolor{mygray}{rgb}{0.5,0.5,0.5}
\definecolor{mymauve}{rgb}{0.58,0,0.82}

\title{GPU-accelerated Guided Source Separation for Meeting Transcription}
\name{Desh Raj$^1$, Daniel Povey$^2$, Sanjeev Khudanpur$^{1,3}$}
\address{
  $^1$CLSP \& $^3$HLTCOE, Johns Hopkins University, Baltimore, USA; $^2$Xiaomi Corp., Beijing, China}
\email{draj@cs.jhu.edu, dpovey@gmail.com, khudanpur@jhu.edu}

\begin{document}

\maketitle
\begin{abstract}
Guided source separation (GSS) is a type of target-speaker extraction method that relies on pre-computed speaker activities and blind source separation to perform front-end enhancement of overlapped speech signals.
It was first proposed during the CHiME-5 challenge and provided significant improvements over the delay-and-sum beamforming baseline.
Despite its strengths, however, the method has seen limited adoption for meeting transcription benchmarks primarily due to its high computation time.
In this paper, we describe our improved implementation of GSS that leverages the power of modern GPU-based pipelines, including batched processing of frequencies and segments, to provide 300x speed-up over CPU-based inference.
The improved inference time allows us to perform detailed ablation studies over several parameters of the GSS algorithm --- such as context duration, number of channels, and noise class, to name a few.
We provide end-to-end reproducible pipelines for speaker-attributed transcription of popular meeting benchmarks: LibriCSS, AMI, and AliMeeting.
Our code and recipes are publicly available: \url{https://github.com/desh2608/gss}.
\end{abstract}

\section{Introduction}

% Speech separation for meeting transcription
Automatic speech recognition (ASR) for meetings is characterized by overlapping speech and far-field multi-channel audio~\cite{Raj2021IntegrationOS}. Speaker overlaps, in particular, result in severe degradation in transcription accuracy, both as a result of inaccurate detection of overlapping segments~\cite{Boakye2008OverlappedSD,Bullock2020OverlapawareDR}, as well as increased ASR errors on these segments~\cite{Kanda2020SerializedOT,Chen2017TheAO,Yu2020AudiovisualMR}. With the rise of deep neural networks (NNs), there have been several advancements in using NN-based mask estimation methods for speech separation~\cite{Luo2020DualPathRE,Luo2019ConvTasNetSI,Subakan2021AttentionIA}. However, these methods are often limited to fully overlapped synthetic speech, and fail to generalize to real, sparse overlaps that are common in multi-talker meetings~\cite{Cosentino2020LibriMixAO,Zeghidour2021WavesplitES,Menne2019AnalysisOD}. Recently, an alternate formulation of speech separation methods, named continuous speech separation (CSS), targeted specifically for sparse overlaps containing an unknown number of speakers, has been proposed~\cite{Chen2020ContinuousSS,Chen2021ContinuousSS}. 

Despite growing popularity of supervised methods, beamforming of multi-channel signals using unsupervised mask estimation remains a strong baseline for multi-talker ASR~\cite{Scheibler2020FastAS,Saijo2022SpatialLF,Ueda2021LowLO,Ikeshita2019AUF}. Among these, the recently proposed guided source separation (GSS) stands out as a particularly effective approach for handling noisy, overlapping speech using diarization information~\cite{Boeddeker2018FrontendPF,Kanda2019GuidedSS}. The method was first proposed for the CHiME-5 challenge, where it provided relative word error rate (WER) improvement of 21.1\% on the multi-array track using oracle segmentation~\cite{Boeddeker2018FrontendPF}. It was later adopted as the challenge baseline for CHiME-6, and used by the winning systems on both oracle and unsegmented tracks~\cite{Watanabe2020CHiME6CT,Arora2020TheJM,Chen2020ImprovedGS,Medennikov2020TheSS}.

\begin{figure}[t]
    \centering
    \includegraphics[width=\linewidth]{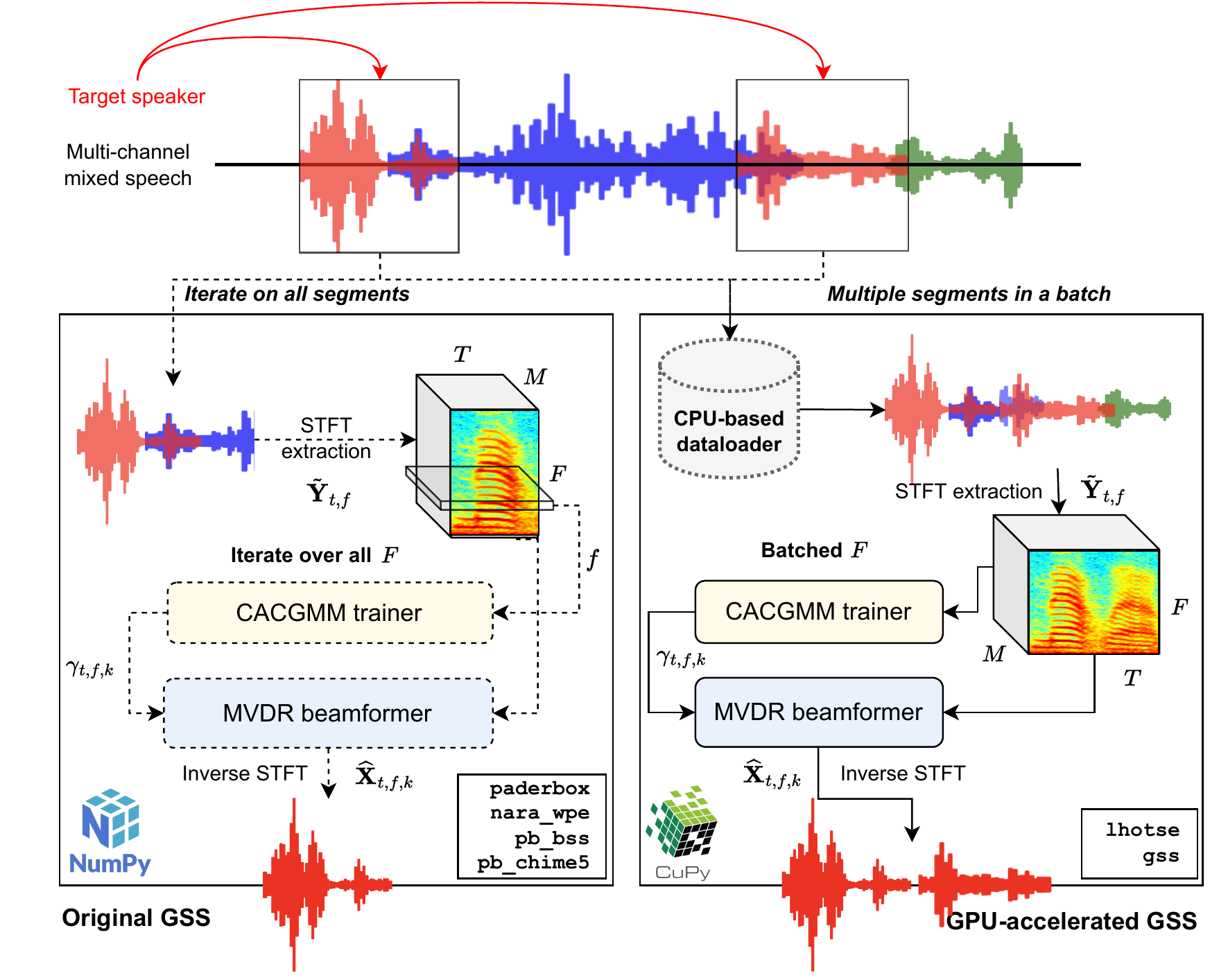}
    \vspace{-1.5em}
    \caption{Overview of batch processing for GPU-accelerated GSS. Solid and dotted lines denote GPU-bound and CPU-bound operations, respectively. The WPE module is not shown.}
    \label{fig:batch}
    \vspace{-0.8em}
\end{figure}

% Advantages of GSS
GSS relies on fundamental ideas from blind source separation (BSS), using spatial mixture models to model the sum of short-time Fourier transform (STFT) bins of multiple speakers~\cite{Comon2010HandbookOB}. It uses diarization information to (i) estimate the number of mixture components, and (ii) avoid the speaker-frequency permutation problem when processing different frequency bins independently. We will describe the algorithm in detail in Section~\ref{sec:gss}. However, despite its strong performance in the CHiME-5 and CHiME-6 challenges, GSS has seen limited adoption in other multi-talker benchmarks, most notably offline meeting transcription, primarily due to its significant computational cost. For instance, enhancing the CHiME-6 \texttt{dev} set using 80 CPU jobs requires approximately 20 hours with the original GSS implementation\footnote{\url{https://github.com/fgnt/pb_chime5}}. There have been some efforts to adapt the offline GSS algorithm for real-time enhancement by relying on limited right context~\cite{Horiguchi2021BlockOnlineGS}, but these are also CPU-bound.

% Contribution
In this paper, we describe our new, publicly-available GPU-accelerated implementation of GSS that aims to remove this computational bottleneck of enhancement. We achieve this primarily by porting all the computations on the GPU, and applying batching at several levels to maximize the GPU memory utilization. Our implementation is inspired by modern deep learning pipelines where background CPU-based workers perform data loading of large tensors, while the data processing is performed by GPUs~\cite{Paszke2019PyTorchAI}. We describe our accelerated implementation in detail in Section~\ref{sec:method}. The resulting 300x speedup allows us to perform ablation experiments using several benchmarks to analyze the importance of factors such as WPE, noise class, context duration, number of BSS iterations, and number of channels, towards GSS performance.

Finally, we provide end-to-end reproducible recipes for meeting transcription of several benchmarks, namely LibriCSS, AMI, and AliMeeting. This includes diarization with and without overlap assignment, GSS-based enhancement, and pretrained models for ASR inference with neural transducers. We believe that our results will provide strong reproducible baselines for all future work on speaker-attributed ASR.

\section{Guided Source Separation}
\label{sec:gss}

\begin{figure}[t]
    \centering
    \includegraphics[width=0.8\linewidth]{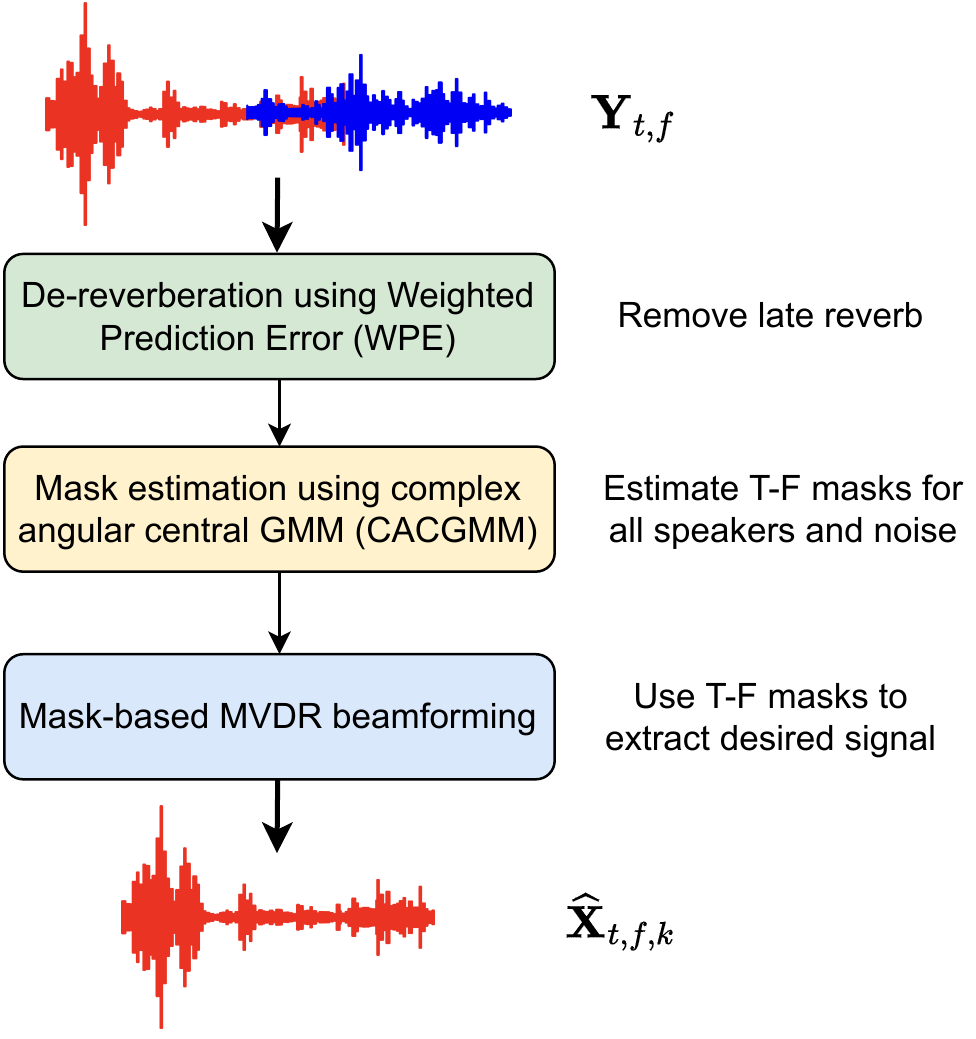}
    \vspace{-0.8em}
    \caption{An overview of the guided source separation (GSS) method for target-speaker extraction.}
    \label{fig:gss}
\end{figure}

We first provide an overview of the GSS algorithm, as proposed in~\cite{Boeddeker2018FrontendPF}. Consider a multi-channel input recording provided in the form of STFT features $\mathbf{Y}_{t,f}\in \mathbb{C}^M$, where $t$ and $f$ are time and frequency bins, respectively, and $M$ is the number of channels. The GSS algorithm assumes the following model of the signal:

\begin{equation}
    \mathbf{Y}_{t,f} = \underbrace{\sum_{k\in K} \mathbf{X}_{t,f,k}^{\mathrm{early}}}_{\mathbf{X}_{t,f}^{\mathrm{early}}} + \underbrace{\sum_{k\in K} \mathbf{X}_{t,f,k}^{\mathrm{tail}}}_{\mathbf{X}_{t,f}^{\mathrm{tail}}} + \mathbf{N}_{t,f},
\end{equation}
where $K$ is the number of speakers in the recording, and ``early'' and ``late'' refer to components of the reverberation. For target-speaker extraction, the objective is to estimate the de-reverberated signal from a desired speaker $k$, i.e., $\widehat{\mathbf{X}}_{t,f,k}$. This estimation is performed in three steps, as described in Fig.~\ref{fig:gss}.

\subsection{De-reverberation using WPE}
\label{sec:wpe}

First, we estimate $\mathbf{X}_{t,f}^{\mathrm{tail}}$, i.e., the ``tail'' part of the reverb, using the popular weighted prediction error (WPE) algorithm~\cite{Nakatani2008BlindSD,Nakatani2010SpeechDB}, and remove it from the signal, followed by normalization to get unit STFT vectors, i.e.,
\begin{equation}
    \Tilde{\mathbf{Y}}_{t,f} = \frac{\mathbf{Y}_{t,f}-\widehat{\mathbf{X}}_{t,f}^{\mathrm{tail}}}{\Vert\mathbf{Y}_{t,f}-\widehat{\mathbf{X}}_{t,f}^{\mathrm{tail}}\Vert}.
\end{equation}

\subsection{Mask estimation using CACGMMs}
\label{sec:mask}

In the second stage, STFT masks are estimated for each speaker (and noise). The mask estimation technique is based on the ``sparsity assumption,'' which assumes that only one speaker is active in each time-frequency bin. Using this assumption, the vector in each T-F bin can be assumed to have been generated from a mixture model where each component of the mixture belongs to a different speaker (or noise class). In the case of GSS, each mixture component is a complex angular central Gaussian (CACG), and hence the mixture model is a CACGMM~\cite{Ito2016ComplexAC}. A CACGMM models sums of unit-normalized complex-valued random variables, and the probability density function at a frequency index $f$ is determined as
\begin{equation}
   p(\tilde{\mathbf{Y}}_{t,f}) = \sum_{k\in K} \pi_{f,k} \mathcal{A}(\tilde{\mathbf{Y}}_{t,f};\mathbf{B}_{f,k}),
\end{equation}
where $\pi_{f,k}$ is the mixture weight of source $k$ at frequency index $f$, and $\mathcal{A}(\mathbf{y};\mathbf{B})$ is a CACG distribution parameterized by $\mathbf{B}\in \mathbb{C}^{M\times M}$ as
\begin{equation}
    \mathcal{A}(\mathbf{y};\mathbf{B}) = \left(\frac{1}{2\pi}\right)^M \frac{(M-1)!}{\lvert\mathbf{B}\rvert} (\mathbf{y}^H \mathbf{B}^{-1} \mathbf{y})^{-M},
\end{equation}
where $(\cdot)^H$ denotes the Hermitian transpose. Mixture model parameters are usually estimated using the EM algorithm that alternates between estimating the state posteriors (in the E-step) and the parameters of the component model (in the M-step). However, there are two problems in applying EM independently for each frequency bin: (i) the number of sources $K$ is unknown; and (ii) the same mixture component may correspond to different sources in different frequency bin. GSS solves both of these problems by assuming that speaker activities are known for the recording, either through an oracle or a diarization system. Given the speaker activities $a_{t,k} \in \{0,1\}$, we convert the time-invariant mixture weights to time-varying weights as
\begin{equation}
    \pi_{t,f,k} = \frac{\pi_{f,k}a_{t,k}}{\sum_{k' \in K}\pi_{f,k'}a_{t,k'}}
\end{equation}
There may still be a permutation problem between the mixture components for the target speaker and the noise signal, since noise is present throughout the recording. To solve this problem, the GSS algorithm adds a ``context window'' to each utterance. We run the EM algorithm on the CACGMM until convergence to obtain the final state posteriors $\gamma_{t,f,k}$ as the estimated speaker masks.

\subsection{Mask-based MVDR beamforming}
\label{sec:mvdr}

Finally, we compute the spatial covariance matrices for the target signal and background signal as
\begin{align}
\Phi_k(f) &= \frac{1}{T}\sum_t \gamma_{t,f,k} \tilde{\mathbf{Y}}_{t,f} \tilde{\mathbf{Y}}_{t,f}^H,\\
\Phi_{\mathrm{bg}}(f) &= \frac{1}{T}\sum_t \left(\sum_{k' \neq k}\gamma_{t,f,k'}\right) \tilde{\mathbf{Y}}_{t,f} \tilde{\mathbf{Y}}_{t,f}^H,
\end{align}
which are then used to compute the minimum-variance distortionless response (MVDR) filter~\cite{Souden2010OnOF,Erdogan2016ImprovedMB} as
\begin{equation}
    \mathbf{h}(f) = \frac{\Phi_\mathrm{bg}^{-1}(f)\Phi_k(f)\mathbf{e}_{\mathrm{ref}}}{\mathrm{tr}\left(\Phi_\mathrm{bg}^{-1}(f)\Phi_k(f)\right)},
\end{equation}
where $\mathbf{e}_{\mathrm{ref}}\in\{0,1\}^M$ is a one-hot vector indicating the reference channel, selected to maximize the signal-to-noise ratio. Finally, the enhanced STFT signal is computed as
\begin{equation}
    \widehat{\mathbf{X}}_{t,f,k} = \mathbf{h}(f)^H \tilde{\mathbf{Y}}_{t,f}.
\end{equation}

\section{GPU-accelerated Inference}
\label{sec:method}

% How we speed up original implementation:
% Batch together all frequency bins instead of iterating one at a time
% Perform einsum matrix operations on GPU
% Fix einsum path for CACG pdf estimation
% Smart batching to utilize GPU memory

The original GSS implementation is slowed down by the following key factors: (i) All the segments are processed sequentially, so processing time for a recording increases linearly with number of identified segments. (ii) A context window (usually 15s) is used for all segments regardless of the segment duration, resulting in a lot of wasted computation for short segments. (iii) For each segment, the CACGMM-based mask estimation is performed by iterating over all frequency bins (usually 513) sequentially. (iv) All computations (i.e., feature extraction, WPE, mask estimation, beamforming, and iSTFT) are performed on the CPU using NumPy~\cite{Harris2020ArrayPW}. A workaround for limitation (i) was provided by using MPI-based multi-processing (or Kaldi-style parallelization~\footnote{\url{https://kaldi-asr.org/doc/queue.html}}) to enhance segments concurrently on a multi-node CPU cluster. Nevertheless, enhancing the CHiME-6 \texttt{dev} set, for instance, may require close to 20 hours (wall clock time) even using 80 CPU jobs (\S~\ref{sec:speed}).  

We propose to accelerate GSS-based inference by leveraging the power of modern GPU hardware and pipelines inspired from neural network training. First, to address limitation (iv), we use CuPy arrays which speed up array operations significantly using CUDA kernels, compared with regular NumPy-based array operations~\cite{cupy_learningsys2017}. Since the most computationally intensive operations in the pipeline (such as CACG probability estimation) involve matrix multiplications (through \texttt{einsum}), GPU-based CUDA kernels are more efficient. However, simply transferring all arrays to CuPy is not sufficient --- for example, limitations (i)--(iii) still require sequential processing, which limits GPU utilization. To maximize GPU utilization and improve real-time factor (RTF), we perform the following additional optimizations.

\begin{enumerate}[wide, labelwidth=!, labelindent=0pt]
\item \textbf{Segment batching.} Instead of processing each segment independently, we batch together multiple segments for inference. However, unlike neural network based training pipelines where batching is performed by stacking sequences in parallel, our batches are formed by concatenating segments sequentially along the time ($T$) axis to create ``super-segments.'' We choose this form of batching because (i) the \texttt{einsum}-based operations are designed to work with 3-D tensors, and (ii) parallel batching of segments with padding would result in wasted memory. Since multiple components of the inference (such as mask estimation and beamforming) compute statistics over the entire segment, we always create super-segments of the same recording with the same target speaker. Furthermore, we only use a single context window for the entire batch (instead of segment-wise context), which further reduces the wasted computations for short segments. This batching technique should work well for the case when optimal reference channels do not vary over the duration of the recording (i.e., when speakers are stationary, which is common for meeting scenarios)\footnote{We also provide the option for using at most one segment per batch, for the case when speakers are not stationary (\S~\ref{sec:speed}).}.

\item \textbf{CPU-based data-loaders.} We ensure that GPU idle time is minimized by off-loading the batch creation process to CPU-based data-loaders (possibly containing multiple workers), similar to deep learning pipelines. Section~\ref{sec:details} provides further details about our Lhotse-based data pipeline.

\item \textbf{Frequency batching.} To address (iii), we modified the CACGMM-based mask estimation to process 3-D tensors $(F,T,M)$ instead of 2-D arrays $(T,M)$. This simple change allows us to process all the frequency bins concurrently in a batch, significantly increasing GPU memory utilization.

\item \textbf{Einsum path optimization.} As mentioned above, several components in the GSS pipeline are implemented using \texttt{einsum}, and uses an optimal path contraction technique to find the path of minimum floating-point operations through the sequence (often resulting in up to 15x speed-up over a naive computation)~\cite{Smith2018opteinsum}. However, the optimal path finding itself is computationally demanding, with a complexity of $\mathcal{O}(N!)$ for $N$ arrays, and since it is performed several times during inference (for example, in each iteration of the CACGMM inference), it overshadows any speed-ups from the actual contracted sum. To remedy this, we cache the optimal computed path in the first iteration and re-use it in subsequent iterations.\footnote{In practice, since our tensor dimensions often have the same relative order across all batches (i.e., $M$<$F$<$T$), we can simply fix the optimal path for all \texttt{einsum} operations. This is because segment batching avoids very short segments that would otherwise result in $T$<$F$.}
\end{enumerate}

\begin{figure}[t]
    \centering
    \includegraphics[width=0.8\linewidth]{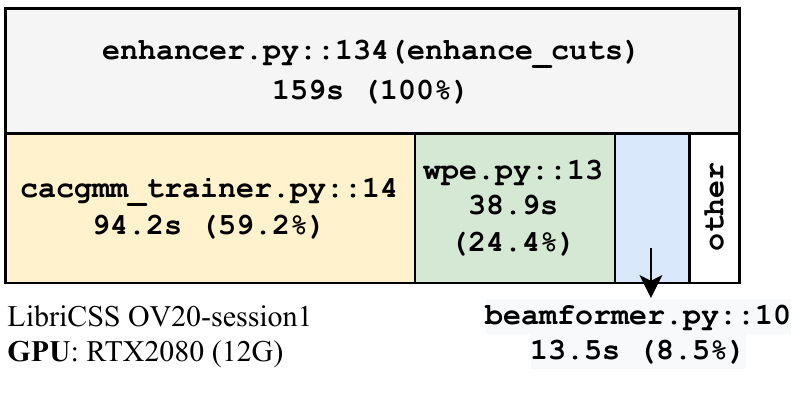}
    \vspace{-1em}
    \caption{Representative output of profiler during enhancement of a single recording. Full stats are available at \href{https://github.com/desh2608/gss/blob/master/test.pstats}{this https url}.}
    \label{fig:profiling}
\end{figure}

Finally, once the enhanced waveform is obtained for the super-segment, we use background worker threads to chunk it into the original segments and save the audios to disk. With all these speed-ups, we were able to enhance a 10-minute LibriCSS recording in 159s (as shown in Fig.~\ref{fig:profiling}), of which mask estimation, WPE, and beamformer constituted 59.2\%, 24.4\%, and 8.5\% processing time, respectively. This is equivalent to a real-time factor (RTF) of approximately 0.3. We anticipate that further speed-ups could be obtained using GPUs with larger memory, by using bigger batches.

\section{Implementation Details}
\label{sec:details}

From an implementation perspective, we can divide the pipeline into two parts. The \textit{data processing} part is tasked with efficiently creating segments and corresponding speaker activities, while the \textit{inference} part performs the actual computations on GPU. We use Lhotse for all data processing, i.e., to store and read recording metadata, to represent speaker activities, and to perform segment batching to create super-segments~\cite{zelasko2021LhotseAS}. For batching we create buckets out of each speaker's segments on-the-fly, and sample speakers in a round-robin manner (see Lhotse's \texttt{DynamicBucketingSampler} and \texttt{RoundRobinSampler}), so that metadata from all segments do not need to be stored in memory. The super-segment obtained from the data-loader is converted to a CuPy array in-place, and all subsequent inference is performed on the GPU. Since we use Lhotse's supervision manifests to store speaker activities, it allows us to use either oracle segments, or read segments from RTTM files (diarization output) with the same data processing pipeline (cf. \S~\ref{sec:asr} and \ref{sec:diar}). A typical recipe for enhancing a corpus is below:

\clearpage

\begin{lstlisting}[language=bash]
#!/bin/bash
# 1. Create Lhotse manifests for corpus
lhotse prepare libricss --type mdm $corpus_dir $data_dir
# 2. Prepare recording-level cuts
lhotse cut simple -r $data_dir/recordings.jsonl.gz -s $data_dir/supervisions.jsonl.gz $exp_dir/cuts.jsonl.gz
# 3. Prepare segment-level cuts
lhotse cut trim-to-supervisions --discard-overlapping $exp_dir/cuts.jsonl.gz $exp_dir/segments.jsonl.gz
# 4. Perform enhancement
gss enhance cuts --max-batch-duration 50.0 $exp_dir/cuts.jsonl.gz $exp_dir/segments.jsonl.gz $exp_dir/enhanced
\end{lstlisting}

\section{Experimental Setup}

\subsection{Data}

\begin{table}[b]
\centering
\caption{Statistics of datasets used for evaluations. LibriCSS does not have a train set. The $k$-speaker durations are in terms of fraction of total speaking time.}
\label{tab:stats}
\adjustbox{max width=\linewidth}{
\begin{tabular}{@{}lrrrrrrrr@{}}
\toprule
\multirow{2}{*}{} & \multicolumn{2}{c}{\textbf{LibriCSS}} & \multicolumn{3}{c}{\textbf{AMI}} & \multicolumn{3}{c}{\textbf{AliMeeting}} \\
\cmidrule(r{2pt}){2-3} \cmidrule(l{2pt}r{2pt}){4-6} \cmidrule(l{2pt}){7-9}
 & \textbf{Dev} & \textbf{Test} & \textbf{Train} & \textbf{Dev} & \textbf{Test} & \textbf{Train} & \textbf{Eval} & \textbf{Test} \\ 
\midrule
\textbf{Duration (h:m)} & 1:00 & 9:05 & 79:23 & 9:40 & 9:03 & 111:21 & 4:12 & 10:46 \\
\textbf{Num. sessions} & 6 & 54 & 133 & 18 & 16 & 209 & 8 & 20 \\
\textbf{Silence (\%)} & 6.2 & 6.7 & 18.1 & 21.5 & 19.6 & 7.11 & 7.7 & 8.0 \\
\textbf{1-speaker (\%)} & 81.3 & 81.2 & 75.5 & 74.3 & 73.0 & 52.5 & 62.1 & 63.4 \\
\textbf{2-speaker (\%)} & 18.6 & 18.5 & 21.1 & 22.2 & 21.0 & 32.8 & 27.6 & 24.9 \\
\textbf{>2-speaker (\%)} & 0.1 & 0.4 & 3.4 & 3.5 & 6.0 & 14.7 & 10.2 & 11.7 \\
\bottomrule
\end{tabular}}
\end{table}

We performed evaluations on three publicly-available meeting datasets: LibriCSS, AMI, and AliMeeting. \textbf{LibriCSS} consists of multi-channel audio recordings of 8-speaker ``simulated conversations'' that were created by combining utterances from the LibriSpeech \texttt{test-clean} set~\cite{Panayotov2015LibrispeechAA}. It comprises 10 one-hour long sessions, each of which is made up of six 10-minute ``mini sessions'' that have different overlap ratios (ranging from 0\% to 40\%). \textbf{AMI} (Augmented Multi-party Interactions) consists of 100 hours of recorded meetings containing 4 or 5 speakers per session~\cite{Carletta2005TheAM}. \textbf{AliMeeting} is a Mandarin-language corpus collected from real meetings, originally designed for ICASSP 2022 M2MeT challenge~\cite{Yu2022M2MetTI}. Each session consists of a 15 to 30-minute
discussion by 2-4 participants. Detailed statistics for all datasets are shown in Table~\ref{tab:stats}.

We used three different mic settings for our experiments: IHM (individual headset microphone), SDM (single distant microphone), and GSS (GSS-enhanced multi-mic). Since LibriCSS does not provide headset recordings, we used the corresponding LibriSpeech utterances concatenated together to simulate IHM. For all datasets, the first channel of the first array was used for the SDM setting. For LibriCSS and AliMeeting, we used all available channels for GSS, whereas for AMI, we used the first of the two arrays. 

\subsection{Models}

We trained separate transducer-based ASR models for each benchmark. For LibriCSS, we used a pretrained Conformer-transducer~\cite{Gulati2020ConformerCT} trained on LibriSpeech. For AMI and AliMeeting, we trained a Zipformer~\cite{zipformer} transducer on a combination of IHM, IHM with simulated reverb, SDM, and GSS-enhanced far-field recordings of the corresponding train set, and the resulting model was used to evaluate all microphone settings. In all cases, we applied three-fold speed perturbation and noise augmentation using MUSAN~\cite{Snyder2015MUSANAM} noises. We used a ``stateless'' decoder consisting of a convolutional layer with a bi-gram context. The model was trained using a pruned RNN-T loss~\cite{Kuang2022PrunedRF} implemented in k2\footnote{\url{https://github.com/k2-fsa/{k2,icefall}}}. For decoding, we used a WFST-based parallel beam search method with beam size 4~\cite{Kang2022FastAP}. Full training recipes and pretrained models are available on Icefall\footnotemark[\value{footnote}].

For the non-oracle segmentation experiments in \S~\ref{sec:diar}, we used a multi-class spectral clustering based diarization system with and without overlap assignment~\cite{Park2020AutoTuningSC, Raj2020MulticlassSC}. The system consists of a Pyannote-based speech activity detector~\cite{Bredin2021EndtoendSS} fine-tuned on the corresponding train set for AMI and AliMeeting. For embedding extraction, we used a pretrained ResNet101-based network~\cite{Landini2020BayesianHC}, which was trained on VoxCeleb~\cite{Nagrani2017VoxCelebAL,Nagrani2020VoxcelebLS} and CN-Celeb~\cite{Fan2020CNCelebAC}. For these experiments, we report diarization error rates (DER) and concatenated minimum-permutation WER (cpWER)~\cite{Watanabe2020CHiME6CT} in order to analyze the impact of diarization errors on downstream ASR. We did not use any collars to compute DERs for LibriCSS and AMI, but a collar of 0.25 was used for AliMeeting following the original work. All diarization recipes, generated RTTM files, and inference pipelines for meeting transcription are publicly available\footnote{\url{https://github.com/desh2608/diarizer}}\footnote{\url{https://github.com/desh2608/icefall/tree/multi_talker}}.

\begin{table}[t]
\centering
\caption{Comparison of close-talk and far-field ASR performance for meeting datasets. The GSS setting uses 7 channels for LibriCSS and 8 channels for AMI and AliMeeting. $^\dagger$LibriCSS IHM refers to the corresponding LibriSpeech utterances. $^\#$For AliMeeting, the numbers are CER.}
\label{tab:asr}
\begin{tabular}{@{}llcccc@{}}
\toprule
\textbf{Dataset} & \textbf{Setting} & \textbf{Ins.} & \textbf{Del.} & \textbf{Sub.} & \textbf{WER} \\ \midrule
\multirow{3}{*}{LibriCSS} & IHM$^\dagger$ & 0.25 & 0.22 & 1.74 & 2.21 \\
 & SDM & 1.06 & 3.12 & 6.59 & 10.77 \\
 & GSS & 0.31 & 0.89 & 2.14 & 3.34 \\
\hline \hline
\multirow{3}{*}{AMI} & IHM & 2.22 & 4.51 & 11.31 & 18.04 \\
 & SDM & 4.01 & 9.59 & 18.50 & 32.10 \\
 & GSS & 2.43 & 6.07 & 14.33 & 22.83 \\
\hline \hline
\multirow{3}{*}{AliMeeting$^\#$} & IHM & 0.97 & 3.78 & 7.32 & 12.07 \\
 & SDM & 1.99 & 10.00 & 14.38 & 26.38 \\
 & GSS & 1.09 & 4.87 & 9.03 & 14.98 \\ 
\bottomrule
\end{tabular}
\end{table}

\section{Results \& Discussion}

\subsection{Far-field ASR}
\label{sec:asr}

We first demonstrate the improvement in far-field ASR performance when using GSS with oracle segmentation, as shown in Table~\ref{tab:asr}. The IHM and SDM settings may be considered as the lower and upper bounds on WER (or CER), respectively. We found that across all the datasets, GSS improved ASR performance significantly, with the \textbf{recovered error rates}\footnote{$\left( W_{\mathrm{SDM}} - W_{\mathrm{GSS}} \right)/ \left( W_{\mathrm{SDM}} - W_{\mathrm{IHM}} \right)$} \textbf{being 86.8\%, 65.9\%, and 80.4\%} for LibriCSS, AMI, and AliMeeting, respectively. As expected, most of the improvement was obtained from recovered deletion and substitution errors, possibly from better recogntion of overlapped speech segments.

\subsection{Effect of diarization}
\label{sec:diar}

\begin{table}[t]
\centering
\caption{Effect of GSS-based enhancement on unsegmented speaker-attributed ASR performance, measured by cpWER (\%). We compare performances using a spectral clustering based diarizer, with and without overlap assignment (OVL). \xmark and \cmark correspond to the SDM and GSS settings from Table~\ref{tab:asr}, respectively.}
\label{tab:diar}
\adjustbox{max width=\linewidth}{
\begin{tabular}{@{}l@{}lccccccccc@{}}
\toprule
\multirow{2}{*}{} & \multicolumn{1}{@{}l}{\multirow{2}{*}{\textbf{Diarizer}}} & \multicolumn{4}{c}{\textbf{DER}} & \multicolumn{1}{c}{\multirow{2}{*}{\textbf{GSS}}} & \multicolumn{4}{c}{\textbf{cpWER}} \\
\cmidrule(r{2pt}){3-6} \cmidrule(l{2pt}){8-11}
& \multicolumn{1}{c}{} & \multicolumn{1}{c}{\textbf{FA}} & \multicolumn{1}{c}{\textbf{MS}} & \multicolumn{1}{c}{\textbf{Conf.}} & \multicolumn{1}{c}{\textbf{Total}} & \multicolumn{1}{c}{} & \multicolumn{1}{c}{\textbf{Ins.}} & \multicolumn{1}{c}{\textbf{Del.}} & \multicolumn{1}{c}{\textbf{Sub.}} & \multicolumn{1}{c}{\textbf{Total}} \\
\midrule
% LibriCSS
\multirow{4}{*}{}
\raisebox{\dimexpr -\totalheight +0.5ex\relax}[0pt][0pt]{\rotatebox{90}{\footnotesize LibriCSS}}\ldelim\{{4}{0.4cm} & 
\multirow{2}{*}{\textbf{Spectral}} & \multirow{2}{*}{1.19} & \multirow{2}{*}{10.37} & \multirow{2}{*}{3.37} & \multirow{2}{*}{14.93} & \xmark & 1.00 & 13.62 & 3.69 & 18.30 \\
&  &  &  &  &  & \cmark & 0.73 & 12.33 & 2.80 & 15.86 \\
\cline{2-11}
& \multirow{2}{*}{\textbf{~~ + OVL}} & \multirow{2}{*}{2.22} & \multirow{2}{*}{3.79} & \multirow{2}{*}{5.33} & \multirow{2}{*}{11.34} & \xmark & 2.64 & 8.09 & 6.36 & 17.09 \\
& &  &  &  &  & \cmark & 1.62 & 7.13 & 3.38 & \textbf{12.12} \\
\hline \hline
% AMI
\multirow{4}{*}{}
\raisebox{\dimexpr -1.5\totalheight +0.5ex\relax}[0pt][0pt]{\rotatebox{90}{\footnotesize AMI}}\ldelim\{{4}{0.4cm} &
\multirow{2}{*}{\textbf{Spectral}} & \multirow{2}{*}{3.24} & \multirow{2}{*}{18.15} & \multirow{2}{*}{4.14} & \multirow{2}{*}{25.53} & \xmark & 2.64 & 20.32 & 15.49 & 38.45 \\
&  &  &  &  &  & \cmark & 2.59 & 18.00 & 12.96 & 33.55 \\
\cline{2-11}
& \multirow{2}{*}{\textbf{~~ + OVL}} & \multirow{2}{*}{7.39} & \multirow{2}{*}{9.63} & \multirow{2}{*}{6.67} & \multirow{2}{*}{23.69} & \xmark & 4.37 & 14.47 & 19.70 & 38.54 \\
&  &  &  &  &  & \cmark & 3.57 & 12.24 & 15.21 & \textbf{31.02} \\
\hline \hline
% AliMeeting
\multirow{4}{*}{}
\raisebox{\dimexpr -0.95\totalheight +0.5ex\relax}[0pt][0pt]{\rotatebox{90}{\footnotesize AliMeeting}}\ldelim\{{4}{0.4cm} &
\multirow{2}{*}{\textbf{Spectral}} & \multirow{2}{*}{0.17} & \multirow{2}{*}{13.60} & \multirow{2}{*}{2.60} & \multirow{2}{*}{16.37} & \xmark & 1.16 & 26.38 & 10.05 & 37.59 \\
&  &  &  &  &  & \cmark & 0.86 & 24.34 & 7.23 & 32.43 \\
\cline{2-11}
& \multirow{2}{*}{\textbf{~~ + OVL}} & \multirow{2}{*}{2.83} & \multirow{2}{*}{5.96} & \multirow{2}{*}{5.64} & \multirow{2}{*}{14.43} & \xmark & 2.32 & 18.82 & 14.30 & 35.44 \\
&  &  &  &  &  & \cmark & 1.69 & 17.00 & 9.76 & \textbf{28.45} \\
\bottomrule
\end{tabular}}
\end{table}

For meeting transcription, it may be hard to obtain oracle segmentation, and often a diarization system is used as a pre-processing step for ASR. In Table~\ref{tab:diar}, we investigate the impact of using non-oracle segmentation with GSS-based enhancement. We found that \textbf{when no enhancement is performed, overlap detection results in little to no cpWER improvement}, since the ASR system is unable to handle overlapping segments. This finding corroborates the results of the winning CHiME-6 system~\cite{Medennikov2020TheSS}, which was able to substantially improve ASR performance on unsegmented recordings using TS-VAD based diarization~\cite{Medennikov2020TargetSpeakerVA}. Using GSS results in significant improvements, with relative cpWER (or cpCER) reductions of 29.1\%, 19.5\%, and 19.7\% on LibriCSS, AMI, and AliMeeting, respectively. 

\subsection{Which factors are most important for GSS?}

\begin{figure}[t]
\begin{subfigure}{0.49\linewidth}
\centering
\includegraphics[width=\linewidth]{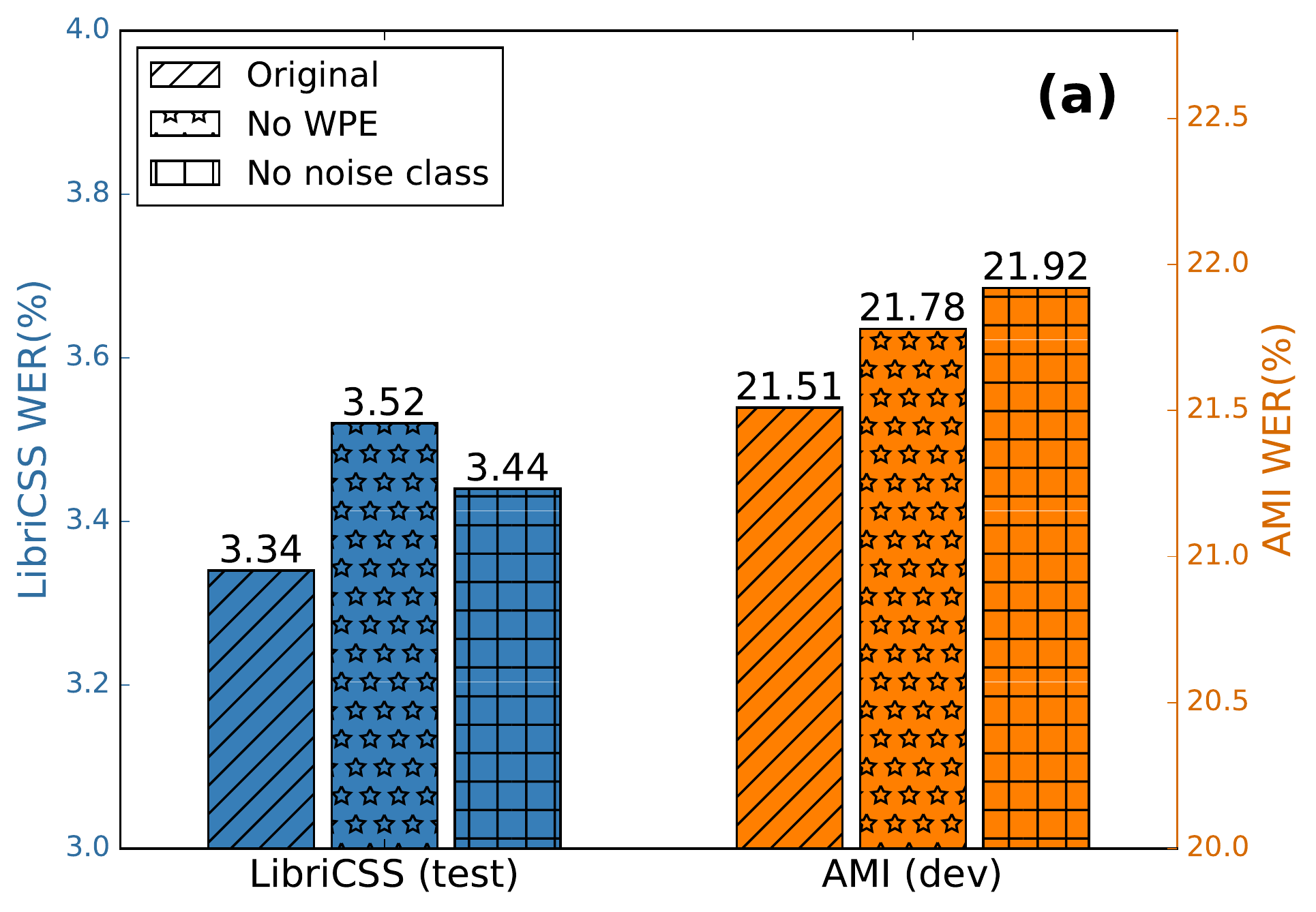}
\captionlistentry{}
\label{fig:wpe}
\end{subfigure}
\begin{subfigure}{0.49\linewidth}
\centering
\includegraphics[width=\linewidth]{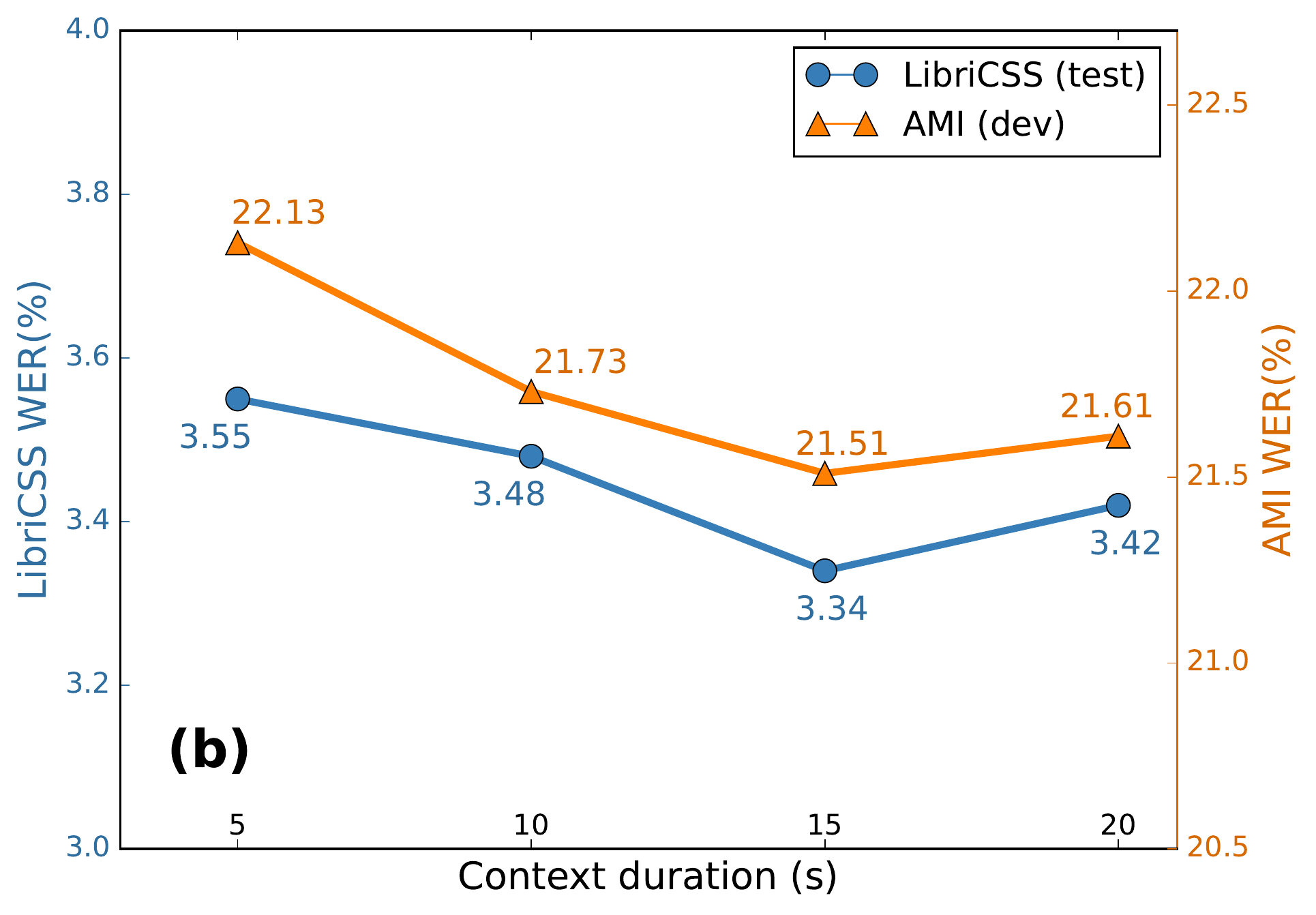}
\captionlistentry{}
\label{fig:context}
\end{subfigure}\hfill
\begin{subfigure}{0.49\linewidth}
\centering
\includegraphics[width=\linewidth]{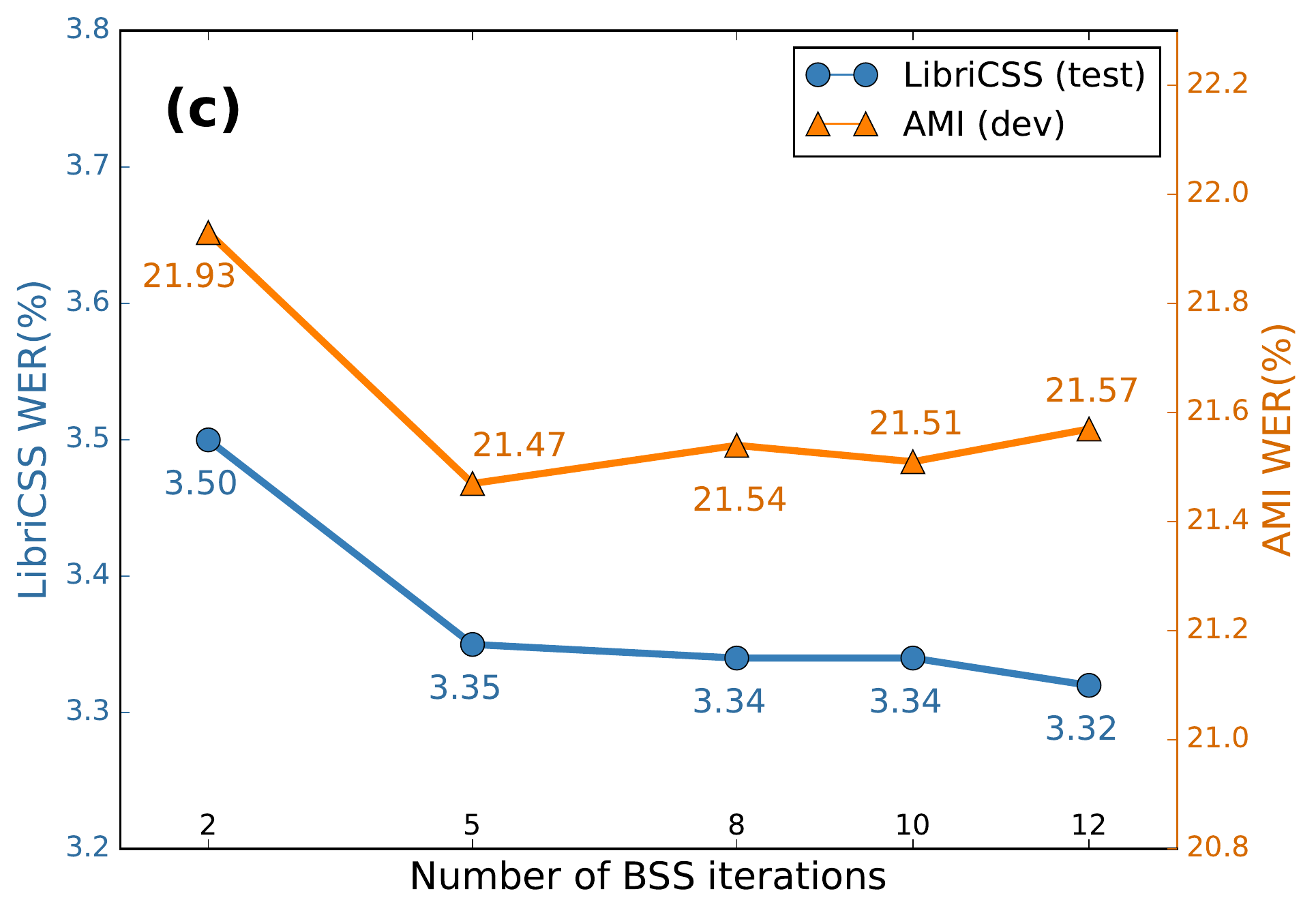}
\captionlistentry{}
\label{fig:iter}
\end{subfigure}
\begin{subfigure}{0.49\linewidth}
\centering
\includegraphics[width=\linewidth]{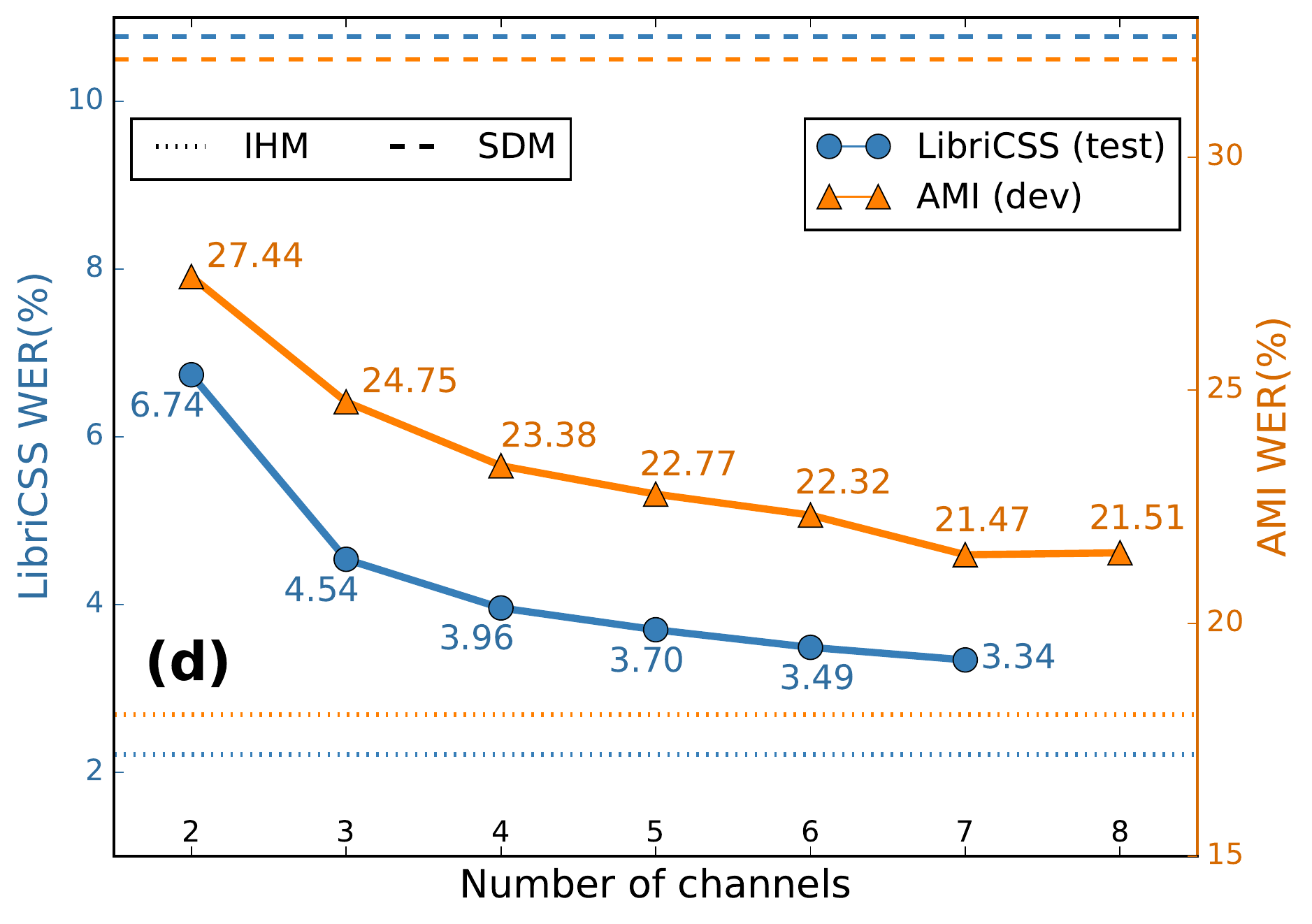}
\captionlistentry{}
\label{fig:channels}
\end{subfigure}\hfill
\caption{Impact of several factors on ASR performance. In each figure, the left and right y-axes denoten WERs for LibriCSS (test) and AMI (dev), respectively, with the axes scaled according to the range of corresponding WER values.}
\label{fig:analysis}
\end{figure}

% Effect of WPE-based dereverberation
% Effect of noise class
% Effect of context duration
% Effect of number of channels
% Effect of number of BSS iterations
We performed ablation studies to investigate the effect of several GSS parameters --- WPE, noise class, context duration, number of iterations for CACGMM inference, and number of input channels --- on the downstream ASR performance, as shown in Fig.~\ref{fig:analysis}. WPE was found to be more important for LibriCSS, while using an additional noise class was more important for AMI (Fig.~\ref{fig:wpe}). This may be because AMI contains occassional background noise, which is absent in LibriCSS. Increasing the context duration from 5s to 15s resulted in consistent WER gains, but adding further context degraded WER (Fig.~\ref{fig:context}). This may be because of inclusion of the target speaker segments in the context if it is expanded too far. A similar observation was made earlier for CHiME-5~\cite{Barker2015TheT}, where a 15s context resulted in better WER compared to a 2s context~\cite{Boeddeker2018FrontendPF}.

For both datasets, increasing the number of BSS iterations (for CACGMM inference) beyond 5 did not result in any WER improvements (Fig.~\ref{fig:iter}). Finally, \textbf{using more input channels was found to be the single most important factor} for better WER performance. For example, using seven input channels resulted in relative WER reduction of 50.4\% and 21.8\% on LibriCSS and AMI, respectively, compared to using two channels. Nevertheless, it follows the law of diminishing returns, as evident by the exponential decay in Fig.~\ref{fig:channels}.

\begin{table}[t]
\centering
\caption{Compute time for our GSS implementation compared with original on CHiME-6 dev set, using all available channels, 15s context, and 20 BSS iterations. ``Time'' is the actual wall clock time (in hours), while ``cum. time'' is the effective total time for all jobs. Speedup is the ratio of the cumulative times.}
\label{tab:speed}
\adjustbox{max width=\linewidth}{
\begin{tabular}{@{}lcrrrrr@{}}
\toprule
\textbf{GSS} & \textbf{Compute} & \textbf{Time} & \textbf{Cum. time} & \textbf{Speedup} & \textbf{WER} & \textbf{+RNNLM} \\
\midrule
\textbf{Original} & 80 x Xeon & 19.3 & 1542.6 & 1.0 & 44.7 & 43.5 \\
\textbf{Ours} & 4 x V100 & 1.3 & 5.3 & 292.2 & 44.2 & 43.1  \\
% \textbf{Ours} & 10s & 5 & RTX (24G) x 4 & 2.33 & 9.31 &  & \\
\bottomrule
\end{tabular}
}
\end{table}

\subsection{Analysis of speed-up}
\label{sec:speed}
% Speed-up compared to original implementation (on CHiME-6?)

We compared our GSS implementation with the original GSS on the CHiME-6 development set in terms of wall clock time and ASR performance, as shown in Table~\ref{tab:speed}. For ASR inference, we used the publicly available Kaldi recipe and pretrained models from JHU-CLSP's submission to the CHiME-6 challenge\footnote{\url{https://github.com/kaldi-asr/kaldi/blob/master/egs/chime6/s5b_track1}}\footnote{\url{https://kaldi-asr.org/models/m12}}~\cite{Arora2020TheJM}. We found that our implementation obtained an \textbf{effective speed-up of 292.2} without any degradation in WER. Since CHiME-6 has non-stationary speakers, we disabled segment batching for this experiment. We can obtain even further speed-ups by enabling this for meeting-like data where speakers are stationary.

\section{Conclusion}

We described our GPU-accelerated implementation of GSS-based front-end enhancement for meeting transcription. On the CHiME-6 benchmark, it was found to be 300x faster than the original implementation, thus removing the computational bottleneck associated with this technique. Through experiments conducted on LibriCSS, AMI, and AliMeeting, we showed that GSS-based enhancement can recover up to 80\% of the WER difference, in going from close-talk to far-field conditions. We also performed several ablation studies to study the effect of GSS parameters, and showed that using more input channels is the single most important factor for better ASR performance. Our \texttt{pip}-installable package is publicly available at \url{https://github.com/desh2608/gss}.

\section{Acknowledgments}

This project was partially funded by NSF CCRI Grant No. 2120435, and a fellowship from Amazon via the JHU-Amazon Initiative for Interactive Artificial Intelligence (AI2AI).

\clearpage

\small
\bibliographystyle{IEEEtran}
\bibliography{main}

\end{document}